\documentclass[11pt]{article}

\usepackage{abstract, adjustbox, algorithm, algpseudocode, amsmath, amsthm, amssymb, appendix, bm, booktabs, caption, etoolbox, lettrine, longtable, listofitems, listings, lmodern, fancyhdr, graphicx, multirow, neuralnetwork, subfigure, tabularray, tabularx, titlesec, titling, wrapfig, xcolor, xurl}
\usepackage[subfigure]{tocloft}
\usepackage[colorlinks=true, allcolors=blue]{hyperref}
\usepackage[english]{babel}
\usepackage{ebgaramond}
\usepackage[scaled]{helvet}
\usepackage[T1]{fontenc}

\usepackage[margin=0.7in, top=1.2in, bottom=1in]{geometry}

\definecolor{box_color}{HTML}{ECF3F6}
\definecolor{box_color_dark}{HTML}{D0E6F0} 
\definecolor{linkcolor}{HTML}{1B6EA2}

\hypersetup{colorlinks=true,linkcolor=linkcolor,urlcolor=linkcolor,citecolor=linkcolor,allcolors=linkcolor}
\algnewcommand{\Inputs}[1]{%
  \State \textbf{Inputs:}
  \Statex \hspace*{\algorithmicindent}\parbox[t]{.8\linewidth}{\raggedright #1}
}

\pretitle{
    \begin{flushleft}
    \vspace{-15mm} 
    \Huge
    \bfseries 
    \sffamily
} 
\posttitle{
    \end{flushleft}
    \vspace{3mm}
    \hrule
    \vspace{3mm}
} 
\preauthor{\begin{flushleft}\large\sffamily}
\postauthor{\end{flushleft}}

\renewenvironment{abstract}
 {\par\noindent\textbf{\sffamily \abstractname} \ \ignorespaces}
 {\par\medskip}
 
\renewcommand\thesection{\Roman{section}} 
\renewcommand\thesubsection{\roman{subsection}} 
\titleformat{\section}[block]{\Large\sffamily\bfseries}{\thesection.}{1em}{} 
\titleformat{\subsection}[block]{\large\bfseries\sffamily}{\thesubsection.}{1em}{} 

\DeclareCaptionFormat{caption_format}{\normalfont\sffamily\fontsize{9}{11}\selectfont#1#2#3}
\adjustboxset{margin=10pt,bgcolor={HTML}{ECF3F6},rndcorners=6}

\DeclareRobustCommand{\tablefont}{%
        \fontencoding{\encodingdefault}%
        \fontseries{m}
        \fontshape{n}
        \fontfamily{phv}
        \fontsize{9}{12}
        \selectfont}
\DeclareTextFontCommand{\texttable}{\tablefont}

\newcommand{\dropcap}[1]{\lettrine[lines=2,lraise=0.05,findent=0.1em, nindent=0em]{{\sffamily{#1}}}{}} 

\usepackage{csquotes}
\usepackage[backend=biber,style=nature,intitle=true]{biblatex}
\AtEveryBibitem{%
  \ifboolexpr{
    test {\ifentrytype{article}} 
    or
    test {\ifentrytype{book}}
  }{%
    \clearfield{url}%
    \clearfield{urldate}%
    \clearfield{urlyear}
    \clearfield{month}
    \clearfield{issn}
  }{}%
}
\usepackage{ragged2e}
\AtBeginBibliography{\RaggedRight}

\title{Mapping the Reddit Bot Ecosystem: Taxonomy and Evolution} 

\date{}

\author{
	\bfseries{Qiusi Sun}$^{\text{1}\star}$,
     \bfseries{Thomas Gaskin$^\text{1}$},
     \bfseries{Branko Blagojevic}$^\text{2}$, \bfseries{Milena Tsvetkova}$^{\text{1,3}\dagger}$
}

\begin{document}

\maketitle

\vspace{-20mm}
{\small\flushleft
\textsuperscript{\textbf{1}}~Department of Methodology, London School of Economics and Political Science, London, United Kingdom; 
\textsuperscript{\textbf{2}}~Pledge Path, New York, United States;
\textsuperscript{\textbf{3}}~Complexity Science Hub, Vienna, Austria

\medskip 
\textit{Corresponding authors:} \quad $^\star$ q.sun27@lse.ac.uk \quad \textsuperscript{$\dagger$} m.tsvetkova@lse.ac.uk
}

\vspace{6mm}

\hrule
\vspace{6mm}

\begin{abstract}
\noindent Automated agents increasingly participate in online communities, yet their population structure and roles remain poorly understood. Using a dataset of 3,389 identified bots and their full activity histories, we construct a taxonomy of bots that post on the news aggregation and social media platform Reddit. Clustering analysis based on temporal, community, linguistic, and semantic features reveals 18 distinct bot ``species'' spanning content-specialized, behavior-driven, and infrastructural roles such as moderation and utility support. In addition, temporal analysis shows that bot numbers and activity expanded rapidly before peaking around the COVID-19 period, then started declining even before Reddit's 2023 API policy changes. However, the overall diversity of bot species has remained remarkably stable. These findings suggest that online bot populations form evolving digital ecosystems.
\noindent

\bigskip

\end{abstract}
\vspace{6mm}
\hrule
\vspace{1mm}
\setcounter{tocdepth}{1}
\tableofcontents
\vspace{4mm}

\newpage

\section{Introduction}
\dropcap{A}rtificial agents such as Twitter/X social bots, financial trading algorithms, and Large Language Model (LLM) chat bots increasingly shape contemporary social and economic systems. They have been implicated in the spread of misinformation \cite{shao2018,vosoughi_spread_2018}, the precipitation of flash market crashes \cite{Kirilenko_2017}, and the disruption of jobs and employment \cite{teutloff_2025}. At the same time, research has demonstrated that they can enhance team communication \cite{traeger_2020}, improve workforce productivity \cite{brynjolfsson_2025}, and support more effective education \cite{Kestin_2025}. More recently, advances in generative AI have enabled artificial agents to not merely assist humans, but also to autonomously interact with one another in emerging AI-native online environments such as Moltbook \cite{jiang_humans_2026}. To understand these transformations, scientists have started viewing human-machine networks and communities through a systemic ecological lens, approaching bots and artificial agents not as tools and artifacts but as social actors with complex behavioral patterns and interactions \cite{rahwan2019,tsvetkova_new_2024}.

This study investigates the diversity and evolution of the bot population of Reddit on the basis of such an ecological perspective. Reddit is one of the most visited news aggregation and social media platforms worldwide, with over 500 million registered users globally as of 2025 \cite{backlinko_team_reddit_2025}. The platform's thousands of topical communities (subreddits) and its commitment to user anonymity and pseudonymity support the deep, asynchronous discussions that set it apart from other major social media platforms such as Facebook, X, and TikTok. Reddit also hosts a large bot population. Similar to other social media platforms, it has bots that operate covertly in coordination with agenda-setting intent \cite{lloyd_there_2025}. However, due to heavy moderation, peer monitoring and voting, covert bots arguably constitute a less severe problem on Reddit. Instead, Reddit's tech-savvy users, relatively open API, and community-developed guidelines such as the ``Bottiquette'' \cite{hurtado2019} facilitate and even encourage the development of helpful bots. These bots serve multiple roles, from helping to maintain the community to enhancing its humorous and geeky culture \cite{massanari2016}. Reddit users often engage with these bots in a performative way, even when their automated nature is clearly disclosed \cite{ma2020}.

Using a large, peer-labeled database of bots on Reddit since 2005, we analyze the accounts' activity to identify distinct bot types. We then apply concepts and operationalizations from evolutionary biology and population ecology to understand the dynamics of this population of non-biological entities: its growth, speciation, internal dynamics, and adaptation to external events. By examining Reddit's bot population at scale, we aim to clarify the roles artificial agents have played within the platform over time, understand how social media platforms may respond to, resist, or incorporate increasingly advanced LLM-based technologies, and lay the groundwork for future research on human-bot interactions.

\section{Background}
To understand the diversity of any population, scientists tend to classify its members into categories. There are two main approaches to classification: typology, which is primarily conceptual and deductive, and taxonomy, which is empirical and inductive \cite{bailey_typologies_1994}.

The literature contains numerous typologies of online bots. One proposed typology distinguishes online bots by intent (benevolent vs. malevolent) and purpose (collect information, execute actions, generate content, emulate humans) \cite{tsvetkova2017a}. Another proposed typology differentiates social media bots by their technological design, purpose, and use \cite{Gorwa_Guilbeault_2020}. The copious literature on Twitter bots suggests additional types: based on the disclosure of automation, it distinguishes between authentic bots \cite{hurtado2019, yang_botometer_2022,yang_botumc_2025} and disguised/camouflaged bots \cite{shevtsov_botartist_2025,yang_botometer_2022}; based on the extent of human involvement, it identifies low-quality scripted bots \cite{shevtsov_botartist_2025}, semi-automated bots/cyborgs \cite{kim_posting_2021,shah_spam_2025}, and paid human agents \cite{hurtado2019}; based on the level of collective coordination, it differentiates between individual bots, network bots \cite{deshmukh_bot_2024}, and vendor-purchased bot squads \cite{caldarelli_role_2020,yang_2020}. Based on the bots' purpose, the same literature identifies administrators/moderators \cite{hurtado2019}, spammers \cite{alipour_lightweight_2025,karatas_sahin_2017,shah_spam_2025,yang_botumc_2025}, content-generators \cite{caldarelli_role_2020,deshmukh_bot_2024,kim_posting_2021},  and amplifiers/astroturfers, which boost the visibility or perceived popularity of content, users, or narratives \cite{caldarelli_role_2020,hurtado2019,karatas_sahin_2017}, as well as impersonators/sybils, which mimic a real identity to gain trust \cite{karatas_sahin_2017,shah_spam_2025}. A typology of Twitch bots similarly proposes categories by purpose for information sharing, moderation and rule enforcement, entertainment, administration (logistics), and community recognition (e.g., welcome and loyalty point distribution) \cite{seering2018}.

While intuitive and common, typologies are inherently constrained by their predefined conceptual dimensions. Researchers must specify in advance which characteristics distinguish categories and arbitrarily choose the number of categories to include. The allocation of cases to categories is mostly manual and theory-driven. As a result, typologies are well suited for organizing existing knowledge but are less effective at revealing previously unrecognized patterns or discovering new species. To circumvent this, taxonomy approaches develop an empirical categorization based on a large set of quantitatively derived features. 

In this work, we propose a taxonomy of bots on Reddit and use it to analyze the bot ecosystem itself. We assemble a dataset of identified bots and analyze their activity to compile a set of features describing its temporal, community, linguistic, and semantic patterns, to which we then apply dimensionality reduction and clustering analysis. This procedure allows us to identify ``species'' -- groups of bots with similar behavior that is distinct from the behavior of other groups. We then study the changes in the species populations over time, driven by internal dynamics and in response to external factors. Specifically, we investigate changes in species diversity, size, and activity, the origin and extinction of bot species, the community structure and ecological niches in the ecosystem, and evidence for convergent evolution, divergent evolution, and adaptation.

Our work is related to two bodies of literature. Methodologically, we borrow from data-driven studies on social media bots that analyze users’ profile information, temporal signatures, linguistic features, and network patterns \cite{chavoshi_debot_2016,chu_2012,cresci_paradigm-shift_2017,davis_2016,dickerson_using_2014,ferrara_2016,hurtado2019,ng_botbuster_2023,peng_unsupervised_2024,pozzana2020,wang_social_2024,yang_botumc_2025}. However, while this corpus often aims to detect covert political misinformation and spam bots that mimic human users, here we aim to differentiate between different types of already identified bots. Conceptually, we mirror the perspective of previous population-level studies of bots on Wikipedia. This research demonstrated that the diversity of the Wikipedia bot ecosystem is key to the platform's resilience to vandalism \cite{geiger2013} and its emergent machine interactivity \cite{geiger_halfaker_2017,tsvetkova2017a}. Here, we similarly focus on the diversity and adaptivity of the Reddit bot ecosystem.

Understanding bot ecosystems extends beyond describing existing forms of automation. As AI agents become increasingly embedded in online platforms, an ecological perspective provides a foundation to study how automated agents emerge, specialize, and collectively shape online communities. Such understanding is also critical for examining how evolving bot ecosystems influence platform governance and affect human-bot and human-human interactions.

\section{Data and Methods}
We compiled a list of Reddit bot accounts from two public crowdsourced bot-ranking projects: botranks.com \cite{mcfarlin_botranks_2023}, which was active in 2020--23, and botrank.net \cite{blagojevic_b0trank_nodate}, which launched in 2018 by building on the data of a similar predecessor and is still ongoing as of 2026. Both projects compile accounts that Reddit users identify as bots and track evaluations of their contributions: users reply ``good bot'' or ``bad bot'' to indicate whether they consider the bot's contribution helpful or unhelpful. \cite{trujillo_when_2021}.

To ensure consistency and reduce false positives, we select accounts with at least ten total votes, yielding 3,389 confirmed bots. Unless otherwise specified, most of our analyses exclude u/AutoModerator, the official moderator bot of Reddit, due to its exceptional activity volume (it has commented more than 600 million times since its creation on July 26, 2015). We note that the datasets contain more ``good bot'' than ``bad bot'' votes and that covert bots identified as ``bad'' may be quick to delete their accounts, rarely crossing the ten-vote threshold we apply universally. Consequently, we are likely undercounting the covert malevolent bots that operate on Reddit. We acknowledge that the analyses here mainly capture relatively long-living, active, and benevolent bots.

For each identified bot account, we retrieve their complete comment and submission histories from the archived Reddit dataset available on Academic Torrents, including the parent posts to retain the conversational structure. The resulting dataset enables both feature-based categorization and interaction-level analysis.

To classify the bots, we compile a set of 33 features describing the temporal, community, linguistic, and semantic patterns of the bots' activity (Table~\ref{tab:features}). Many of the features are proportions or composite indices within the range $[-1, 1]$ or $[0, 1]$; for the rest, which are mainly counts or time intervals, we log-transformed using $\log_{10}{(x+1)}$ and rescaled to $[0, 1]$. We use Principal Component Analysis (PCA) on the correlation matrix to identify and remove highly correlated variables. We ranked features by summing their percentage contributions to the first two principal components, then retained the smallest ranked set whose cumulative share of this combined contribution reached 90\%. This resulted in 23 features accounting for 90.47\% of the total contribution. The retained features comprise: hourly entropy, variance in response time, and mean inter-post time; number of subreddits, subreddit specialization, trigger-based interaction (e.g., command or name invocation), and similarity to parent posts; lexicon size, lexical diversity, and sentiment; and 13 macro domain topic frequencies~(Fig.~\ref{fig:feature-distributions}). For the latter, we applied BERTopic with Sentence-BERT embeddings to identify latent semantic themes in the bot-generated text. The resulting fine-grained topics were then manually grouped into 13 macro domains to capture broader cultural niches.

We apply hierarchical clustering to the retained 23 features to identify distinct bot archetypes for the taxonomy. We compared candidate values of $k$ using the average silhouette score, which measures how well observations match their assigned cluster relative to other clusters. The selected $k = 18$ solution had an average silhouette score of 0.245 and provided the most interpretable clustering~(Fig.~\ref{fig:silhouette-scores}). The number and general composition of the clusters were similar to those obtained with $k$-means clustering. We visualize the clusters with t-SNE (t-distributed stochastic neighbor embedding), which projects each bot as a point onto a two-dimensional plane, where Euclidean distance between points reflects overall similarity in the 23 features. The absolute orientation (angle) of the plane has no intrinsic meaning as t-SNE is rotation-invariant and the embedding is stochastic; only relative distances between points should be interpreted.

We study bot communities and niches by visualizing the networks of co-posting interactions between bots by year. We start with bipartite graphs where edge weights $\tilde{w}_{is}$ indicate the number of posts a bot $j$ contributed to a particular subreddit $s$ in a given year from 2009 to 2025. We create a one-mode projection of these graphs by taking the minimum weight of each edge two bots share with a common subreddit, then summing over all such shared edges, 
\[ 
    w_{ij} = \sum_s \min (\tilde{w}_{is}, \tilde{w}_{js}).
\]
The edge weights in the resulting network can thus be interpreted as the maximum number of encounters (in the form of comments/replies to each other's posts) two bots could have had in a given year.

\begin{figure*}[!t]
\centering
\includegraphics[width=\textwidth]{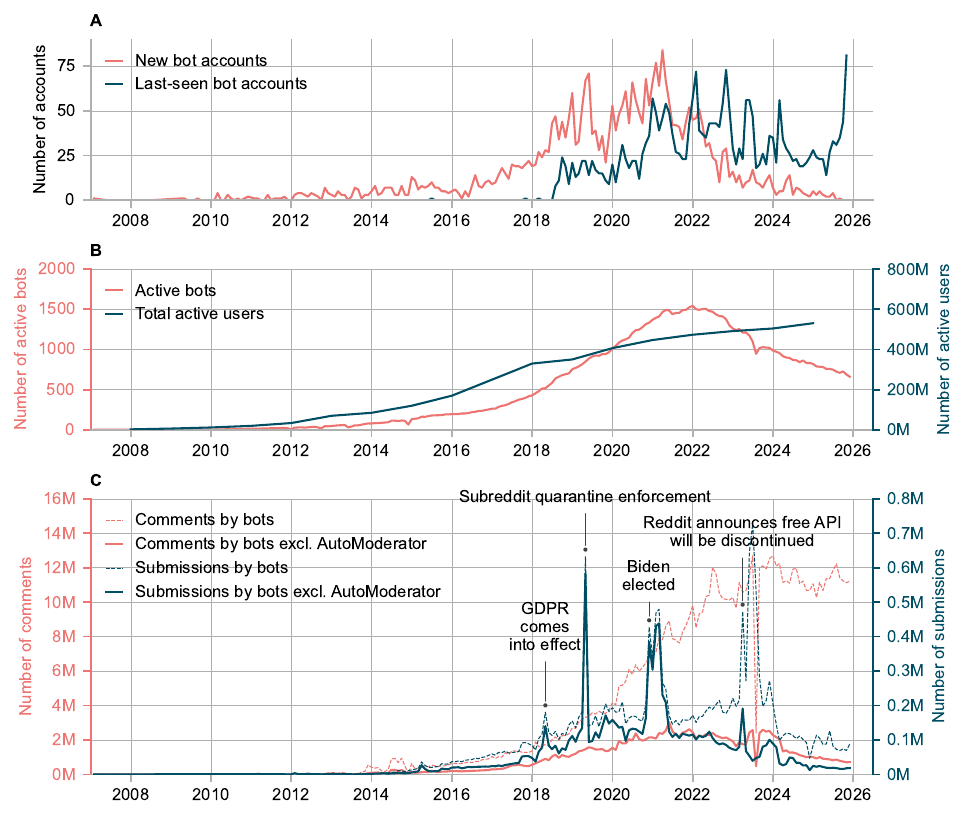}
\caption{Reddit bot accounts and activity over time. A) Number of newly created and last-seen bot accounts. "Last-seen" denoted the date of the final comment or submission captured in our dataset and does not necessarily indicate account deletion. B) Number of active bots and total number of active monthly users based on data from seo.ai \cite{seoai_how_2025}. C) Number of comments and submissions contributed by the bots, with and without Reddit's official moderator bot AutoModerator. The callouts indicate relevant platform updates and external events that could explain the observed discontinuities.}
\label{fig:population}
\end{figure*}

\begin{figure*}[!t]
\centering
\includegraphics[width=\textwidth,height=0.82\textheight,keepaspectratio]{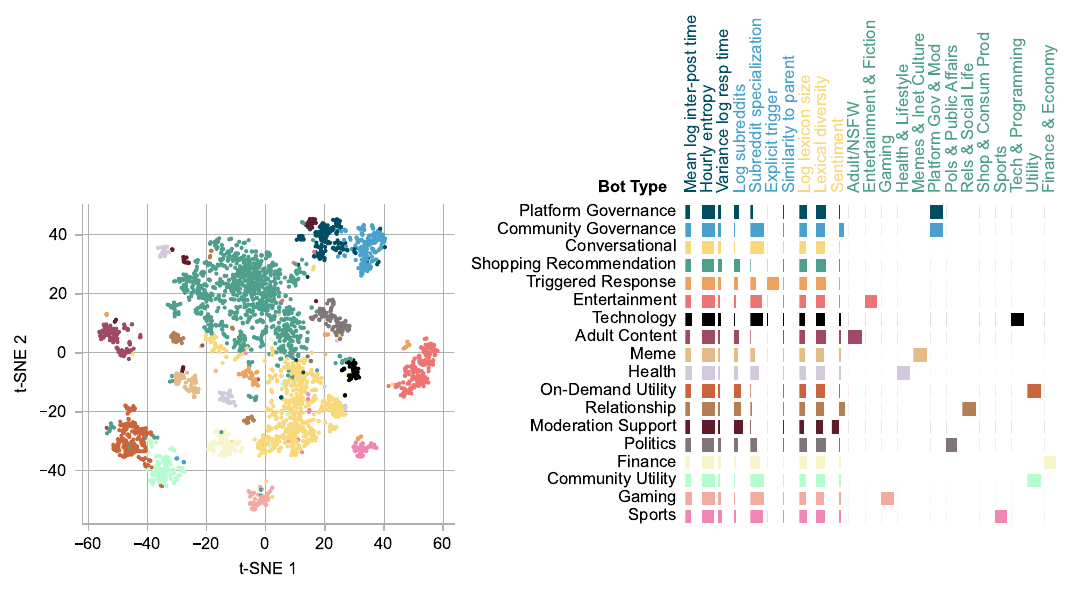}
\caption{t-SNE plot of the 18 bot types identified by the hierarchical clustering, with typical feature values per cluster. The width of the bars in the table to the right corresponds to the median value of the feature for the cluster (values range from $0$ to $1$ for all features except for sentiment, for which they can range from $-1$ to $1$); these results were used to assign the descriptive labels of the clusters. The colors of the bars in the table and the colors of the markers in the t-SNE plot identify clusters. Temporal features are printed in dark blue font, community features in light blue, linguistic features in yellow, and semantic features in green.}
\label{fig:taxonomy}
\end{figure*}

\section{Results}
Reddit launched in June 2005. The earliest bot account in our dataset was created onloy six months later, on December~9,~2005, and the first bot comment in our data was observed on February~5,~2007. This indicates that automation emerged early in the platform's history. Bot account creation accelerated between 2017 and 2021, marking a period of rapid population growth~(Fig.~\ref{fig:population}A). The number of active bot accounts reached its peak during the COVID-19 pandemic, mirroring a broader increase in human activity on the platform, but started declining in the beginning of 2022, even before Reddit's announcement in April 2023 that API access would become paid~(Fig.~\ref{fig:population}B). Bot submissions spiked just before the API changes were implemented on July 1, 2023 and temporarily dropped to nearly zero immediately after~(Fig.~\ref{fig:population}C). Although activity partially rebounded afterward, it continued on a downward trajectory through 2024 and 2025. However, this decline in commenting activity disappears if we account for AutoModerator, which today is responsible for an order of magnitude higher activity on the platform than the rest of the bot population combined. Together, these trends present evidence for a period of initial population growth, followed by a period of population decline, accompanied by the consolidation and centralization of activity by the official Reddit moderation algorithm. This long-term shift, alongside the more short-term dynamics, reveal that the bot population is responsive to both human engagement cycles and structural platform changes.

\begin{figure*}[!t]
\centering
\includegraphics[width=\textwidth]{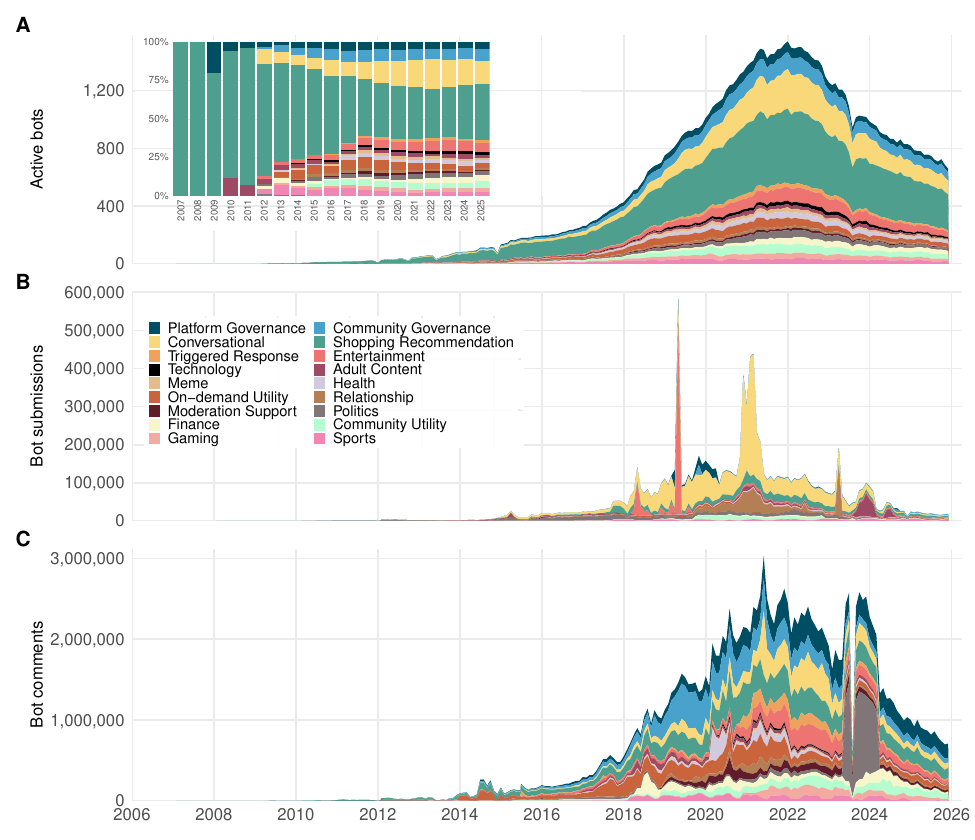}
\caption{Population sizes and activity of the 18 Reddit bot types over time. A) Stacked area chart of active bot accounts, with an inset stacked bar chart showing the annual proportional composition of active bot types. B) Stacked area chart of submissions. C) Stacked area chart of comments.}
\label{fig:activity}
\end{figure*}

Our taxonomy yields 18 bot archetypes that reflect both content specialization and functional differentiation within the bot ecosystem~(Fig.~\ref{fig:taxonomy}, Fig.~\ref{fig:dendrogram}). Several bot types are defined by content domains, such as Technology and Programming, Gaming, and Politics and Public Affairs, indicating clear semantic niches. Other types are primarily distinguished by behavioral patterns, including subreddit specialization, trigger dependence (whether activities are explicitly triggered by users), and negative sentiment, suggesting differences in posting rhythm and interaction style. We also identified four distinct bot types associated with governance and moderation activities, as well as utility-oriented functions, reflecting specialized roles within the Reddit ecosystem. Tracing the population sizes of the 18 identified bot types reveals that the diversity of the bot ecosystem is remarkably persistent and resilient. Most bot types appeared by 2016 and none of them have become extinct as of 2026. In fact, the relative proportion of the different bot types seems remarkably stable despite changes in the total population size~(see inset in Fig.~\ref{fig:activity}A). As a robustness check, we repeated the clustering using behavioral and linguistic features separately. Neither feature subset revealed additional interpretable structure beyond the combined feature set, supporting our integrated approach.

The activity of the different bot types exhibits more distinct temporal patterns~(Fig.~\ref{fig:activity}B, C). Bots associated with governance, moderation, and utility functions maintained a relatively stable presence over time but other bot types experienced distinct high-activity periods and lulls. For example, Conversational bots greatly increased their submission activity at the end of 2020 and beginning of 2021: these were GPT2 models trained on specific subreddits that ``conversed'' e.g. on r/SubSimulatorGPT2, r/SubSimGPT2Interactive, as well as some niche personal projects or localized experimental subreddits~(Fig.~\ref{fig:activity}B). Entertainment bots exhibited two brief spikes in May 2018 and May 2019. The May 2018 spike was largely driven by u/Roboragi, a widely used anime information bot active across anime communities, whereas the May 2019 spike was primarily attributable to u/quzingler\_bot, which generated Shakespeare-inspired dialogue~(Fig.~\ref{fig:activity}B). Relationship bots had a moment in April 2023 when u/rlcd-bot, an identity verification service promoting user trust, came alive, and u/VisualMod, an AI-powered virtual community character, became popular~(Fig.~\ref{fig:activity}B). While the former supported identity-based social interactions, the latter's increased prominence coincided with the widespread adoption of generative AI, suggesting that advances in AI technology may have reshaped the roles and visibility of socially interactive bots. Likewise, Adult Content bots posted actively at the end of 2023 when automated promotion was widespread across Reddit's NSFW (not safe for work) creator communities~(Fig.~\ref{fig:activity}B). Community Governance bots increased their commenting activity in 2019 but have since been largely replaced by broader Platform Governance support bots, on top of the general AutoModerator takeover (Fig.~\ref{fig:activity}C). Politics bots increased their commenting shortly before the API changes in July 2023 and in the months afterwards. During that period, they fled to NSFW subreddits such as r/TrueFMK (which was banned later by Reddit for hosting misogynistic content) and facilitated community participation in Reddit's API protest by encouraging users to pause submissions and votes. Today, these bot types include fewer active accounts than at their historical peaks, reflecting the overall contraction of the Reddit bot ecosystem rather than the disappearance of particular species.

\begin{figure*}[!t]
\centering
\includegraphics[width=\textwidth]{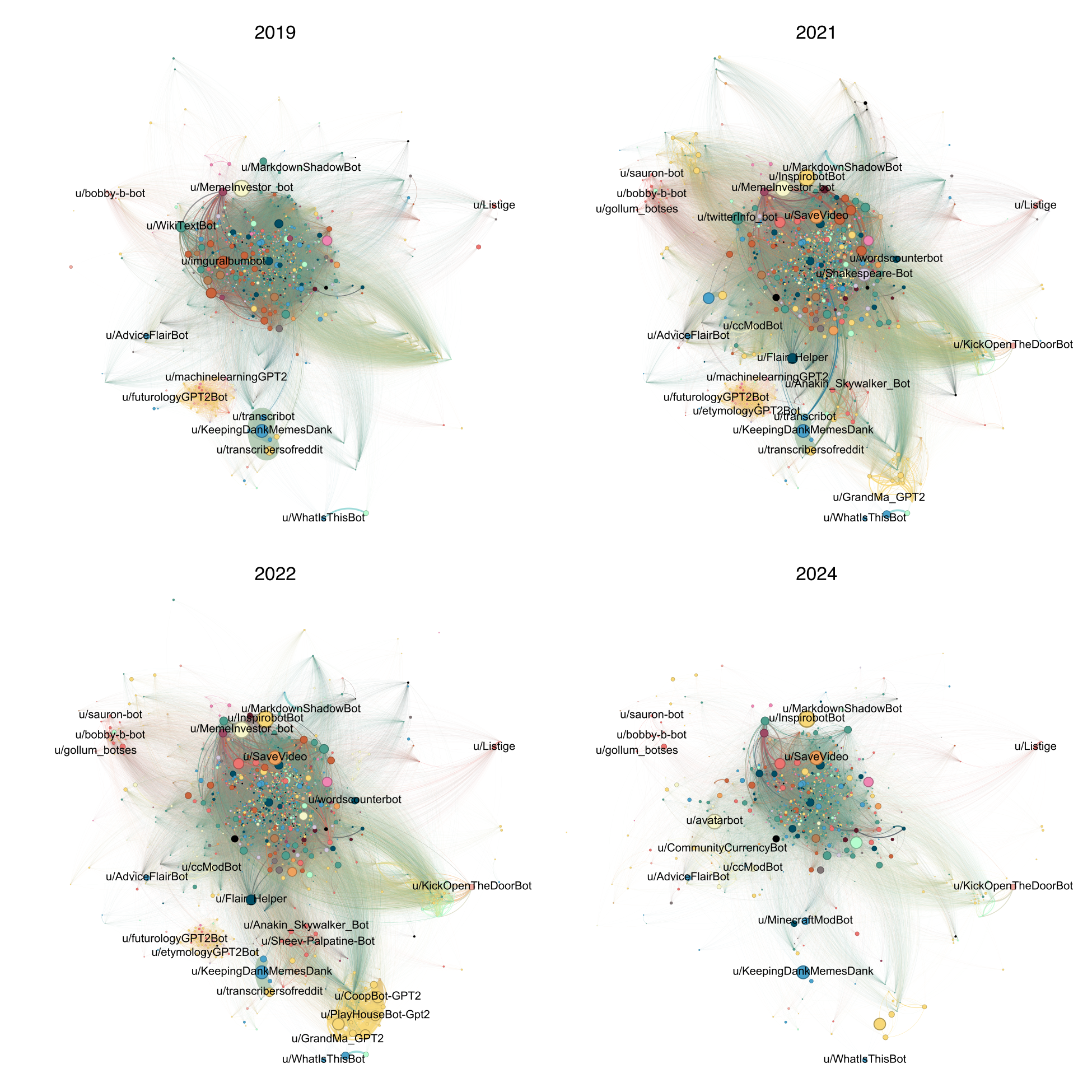}
\caption{Co-posting networks of Reddit bot types in 2019, 2021, 2022, and 2024. Edge weights represent the shared number of posts in a subreddit, summed over all common subreddits to which the two bots posted. Node sizes correspond to the total number of posts made on the platform in that year and node colors map onto the bot types from Figs.~\ref{fig:taxonomy} and \ref{fig:activity}. The networks exclude AutoModerator. The networks are drawn in Gephi with the OpenOrd layout, which is a fast algorithm particularly suitable for identifying clusters in large undirected weighted networks.}
\label{fig:networks}
\end{figure*}

Most bots post across multiple subreddits and are thus tightly interconnected in one large cluster~(Fig.~\ref{fig:networks}). At the peak of the Reddit bot population size and activity in 2021, there were additionally several smaller and well-defined communities, notably, four communities of GPT2 bots, three of which served as simulated environments where bots simulate Reddit users across many everyday topics (r/SubSimulatorGPT2, r/SubSimGPT2Interactive, and r/CoopandPabloPlayHouse), and a technical showcase community for generative models (r/CoopandPabloArtHouse). In 2024, when the population had dramatically decreased, these communities had largely dissolved and only one simulated environment that involved human users (r/SubSimGPT2Interactive) had grown.

\section{Discussion}
The study provides, to our knowledge, the first population-level characterization of community-recognized Reddit bots, tracing the diversity and evolution of the Reddit bot ecosystem from 2005 to 2025. The findings reveal substantial diversity in the bots' semantic, functional and behavioral characteristics, alongside a dramatic decline in the bot population in recent years. Although this contraction is commonly associated with Reddit's API pricing changes introduced in July 2023, the temporal patterns observed here indicate that the decline was already apparent in 2022, predating the policy change. Despite this dramatic contraction of the Reddit bot population, its species composition remained remarkably persistent throughout the observation period. Although the absolute population and activity of nearly all bot species declined, most ecological roles persisted, and their relative representation changed little after the major species had emerged.

While the population composition remained stable, the distribution of activity became increasingly centralized through a single bot, Reddit's official u/AutoModerator. Introduced as a community-developed moderation bot and later, around 2015, integrated into Reddit as an official platform-supported tool, AutoModerator provided a shared yet highly configurable infrastructure that moderators could adapt to local community needs \cite{jhaver2019}. Our analyses suggest that its growing dominance reflects the institutionalization of routine governance within a common platform-supported framework. AutoModerator has therefore become a layer of locally configured governance that complements specialized community bots. This pattern suggests a growing centralization of automated platform functions. Such centralization may improve consistency and administrative efficiency. However, compared with a more distributed system composed of many independently developed bots, a single, platform-managed system may reduce functional diversity and make the ecosystem more vulnerable to random outages or attacks and less resilient to external shocks or natural fluctuations of needs and behaviors. This shift may also reduce opportunities for community-driven innovation, as independently developed bots can be tailored to the specific norms and evolving needs of individual communities in ways that a platform-wide system cannot easily accommodate.

From an evolutionary perspective, changes in the ecosystem are likely shaped by both endogenous and exogenous processes. Endogenous processes originate within the bot population itself, including the emergence of new variants or subtypes within existing species through modification of existing designs or technological innovation. The temporary emergence of GPT2-based bot communities illustrates how advances in AI technology can generate novel variants within existing ecological roles that may later contract as technological and platform conditions change. Existing functional roles may also diversify through the recombination of existing capabilities, producing variants that integrate features previously associated with distinct ecological roles. Exogenous processes arise from changes in the surrounding environment. Platform interventions, such as Reddit's API policy changes, may act as selective pressures that alter the abundance and activity of different bot species without necessarily altering the overall composition of the ecosystem. At the same time, the bot ecosystem likely coevolves with the broader Reddit environment, as changes in user behavior, community norms, and platform content reshape the demands for automation and shift the relative prominence of existing ecological roles.

We note some caveats and limitations to this study. The study focuses on authentic, overt, and user-identified Reddit bots. It does not capture covert bots, coordinated botnets, or automated accounts that are not publicly recognized as bots. Consequently, the observed decline in the bot population should not be interpreted as evidence that automation on Reddit has necessarily decreased. Advances in large language models may have made automated accounts more difficult for users to recognize, reducing the likelihood that they are identified through community-driven conventions such as the ``good bot/bad bot'' rating. Likewise, changes in user awareness or use of these conventions over time may influence how many and which types of bots are included in the present dataset, although the Botrank data does not indicate any systematic trends to support this (Fig.~\ref{fig:annual-composition}). Furthermore, our study characterizes bots through their observable posting activity. Many forms of automation on Reddit operate without producing public posts, including bots that crawl or scrape content, send private messages, distribute awards, or perform other background functions. These forms of automation are not captured by the present analysis and may represent an important component of the Reddit platform. The taxonomy could also potentially be misused to profile automated accounts or help bot developers design systems that evade identification. We therefore present it as a population-level analytical framework rather than a tool for targeting individual accounts.

The observed contraction of the Reddit bot ecosystem may have broader implications for the future evolution of online automation. A larger and more active population provides greater opportunities for new forms of automation to emerge, coexist, and evolve in response to changing community needs. Continued contraction may therefore limit the ecosystem's capacity for innovation over time and reduce opportunities for community-driven experimentation and the development of specialized automated services. These observations are particularly relevant given the rapid advancement of generative AI. Large language models are expanding the capabilities of automated agents while simultaneously changing the environments in which they operate. Recent reports have raised concerns about the use of LLM-generated accounts in unethical online experiments \cite{ogrady_unethical_2025}, as well as the growing volume of AI-generated content that places an additional burden on users and volunteer moderators \cite{tenbarge_ai_2025}. In response to these technological developments, Reddit has introduced stricter account rules \cite{gupta_reddit_2026}. As a result, public perceptions and responses of automation may also be changing. For instance, Twitter data shows that ``bot shaming'' in replies has increasingly become a form of dehumanization \cite{assenmacher_you_2024}. Although these developments cannot be directly linked to the population-level patterns observed here, they illustrate a rapidly changing technological and governance landscape in which online bot ecosystems continue to evolve, while also highlighting the need for future research on how advances in AI are reshaping social norms, perceptions of authenticity, and interactions between human and automated participants in online communities.

The extent to which our findings generalize beyond Reddit remains an open question. Reddit possesses several characteristics that may shape the composition and evolution of its bot ecosystem, including community-specific norms, publicly accessible user profiles and interaction histories, community voting, extensive volunteer moderation, and relatively active platform interventions. Together, these features create a social and institutional environment in which automated accounts are both more visible and more readily subject to community and platform oversight. These characteristics may also help explain an additional pattern observed in the Reddit bot ecosystem. Rather than disappearing entirely, automation may be shifting from decentralized, community-developed bots toward more institutionalized forms of platform-support automation, as reflected by the increasing dominance of AutoModerator. This pattern echoes the institutionalization of bots documented on Wikipedia, where automation became increasingly integrated into the platform's governance infrastructure \cite{geiger2011}.

Whether similar patterns characterize other social media platforms remains uncertain. Platforms such as Facebook and X have faced more persistent concerns surrounding misinformation bots and the growing volume of AI-generated content, reflecting different governance structures, moderation practices, and platform incentives. The evolutionary trajectory documented here should therefore not be assumed to represent online bot ecosystems more generally. Nevertheless, our findings suggest that the evolution of online automation need not follow popular narratives such as the ``Dead Internet'' theory that emphasize ever-increasing bot populations. On Reddit, the population of independently developed, publicly recognized bots has declined, while automation has become increasingly centralized through platform-supported infrastructure. This shift may represent not a reduction in automation, but a transformation of its role within online communities. Such transformation may involve important trade-offs. Although institutionalized automation can improve consistency, scalability, and governance, the decline of community-developed bots may also reduce the scope for playful, creative, and experimental forms of automation that historically contributed to Reddit's participatory culture. It may become an important challenge for online platforms that increasingly integrate AI into official infrastructure to preserve space for community-driven innovation. At the same time, the advent of platforms such as Moltbook that are designed specifically for interactions among AI agents suggests that new forms of AI-native social ecosystems may emerge outside traditional human-centered online communities.

The present study opens several directions for further research. First, the taxonomy developed here represents one possible characterization of the Reddit bot ecosystem. We combined temporal, community, linguistic, and semantic features to capture multiple dimensions of bot behavior, but alternative taxonomies could emphasize a subset of these dimensions, add other behavioral dimensions, incorporate additional features, or explore different clustering approaches. Such work would help refine our understanding of bot diversity on Reddit and across online communities. Second, although we examined relationships among bots through co-posting networks, we did not investigate direct interactions between bots through comments and replies. Similarly, interactions between humans and publicly recognized bots remain largely unexplored; the taxonomy presented here provides a useful starting point for identifying different bot types and examining how users engage with them across communities. Finally, an important next step is to investigate the coevolution of bots and humans over time. As the Reddit bot ecosystem continues to evolve, are interactions with bots becoming less frequent or changing in nature? At the same time, how are changes in the bot ecosystem and the public perception of automation influencing human behavior and reshaping the patterns of both human-bot and human-human interactions within online communities?

\paragraph*{Acknowledgements}
This work is supported by ERC grant HUMANET No.101170272. We are grateful to Brandon McFarlin for sharing his Botranks data.

\printbibliography

\clearpage\section*{Supplementary Information}
\setcounter{figure}{0}
\renewcommand{\thefigure}{S\arabic{figure}}
\setcounter{table}{0}
\renewcommand{\thetable}{S\arabic{table}}

\IfFileExists{figures/figureS1.pdf}{\begin{figure}[H]\centering\includegraphics[width=\textwidth,height=0.80\textheight,keepaspectratio]{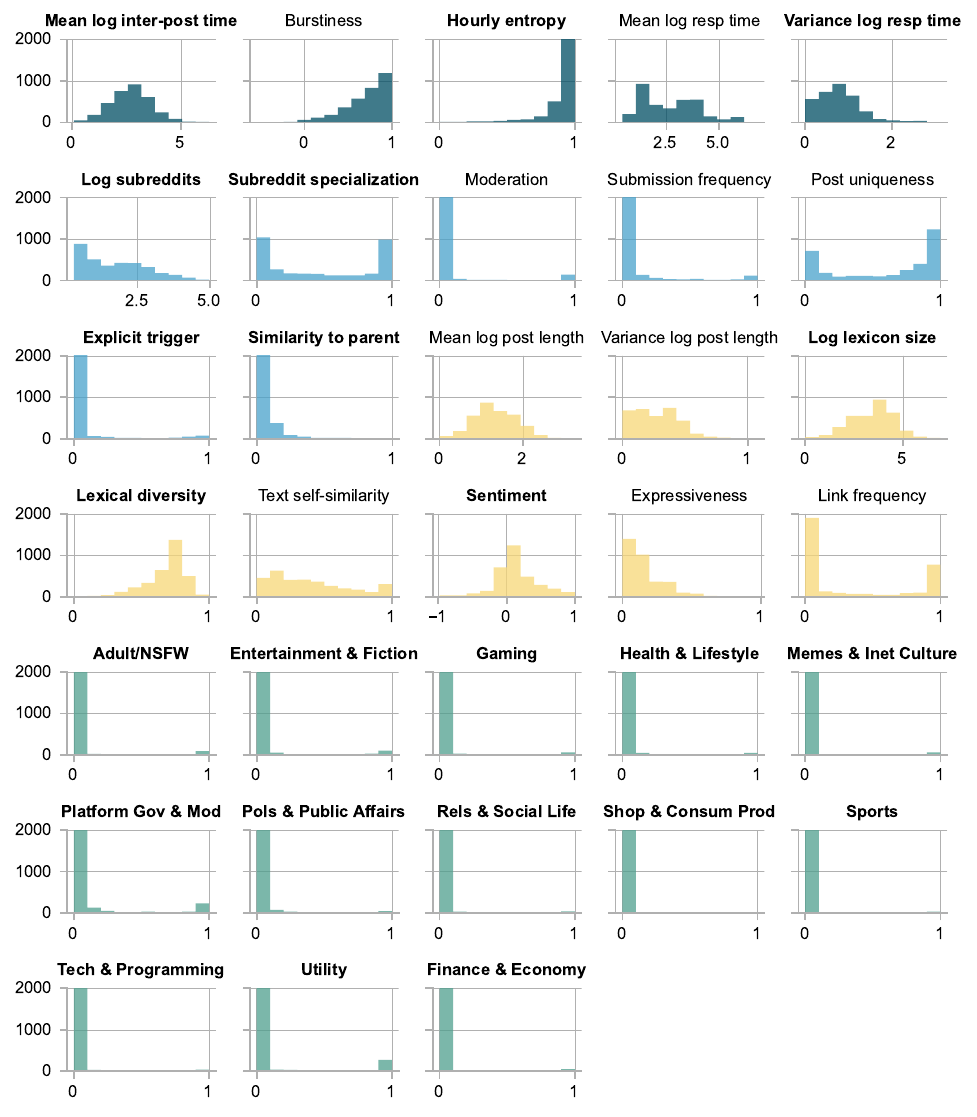}\caption{Distributions of the unscaled features. Temporal features are shown in dark blue, community features in light blue, linguistic features in yellow, and semantic features in green. Bold text highlights the 23 features used in the clustering algorithm. Note that the y-axis is cut off at 2000, which is around 60\% of bots.}\label{fig:feature-distributions}\end{figure}}{\begin{figure}[H]\centering\fbox{\parbox[c][0.28\textheight][c]{0.84\textwidth}{\centering Placeholder: upload figures/figureS1.pdf}}\caption{Distributions of the unscaled features. Temporal features are shown in dark blue, community features in light blue, linguistic features in yellow, and semantic features in green. Bold text highlights the 23 features used in the clustering algorithm. Note that the y-axis is cut off at 2000, which is around 60\% of bots.}\label{fig:feature-distributions}\end{figure}}\clearpage
\begin{figure}[H]
\centering
\includegraphics[width=0.78\textwidth,keepaspectratio]{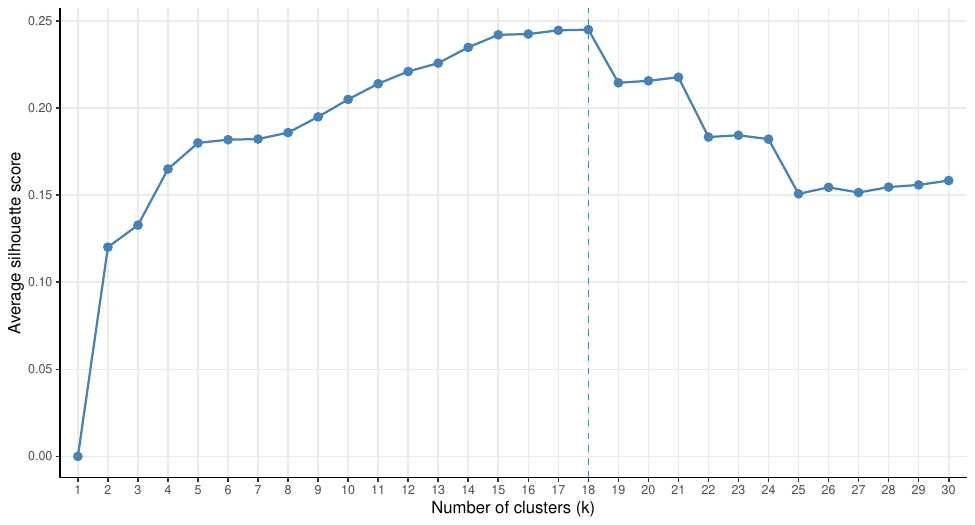}
\caption{Average silhouette scores across candidate numbers of clusters $k$. The selected $k=18$ solution had an average silhouette score of 0.245.}
\label{fig:silhouette-scores}
\end{figure}
\IfFileExists{figures/figureS3.pdf}{\begin{figure}[H]\centering\includegraphics[width=0.78\textwidth,keepaspectratio]{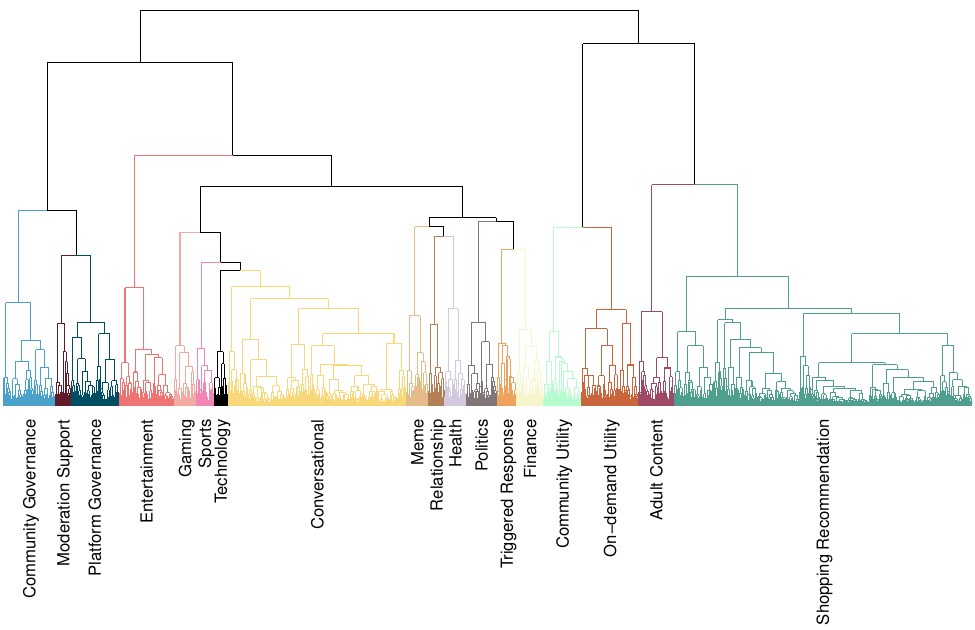}\caption{Dendrogram from the hierarchical clustering used to identify the 18 bot types.}\label{fig:dendrogram}\end{figure}}{\begin{figure}[H]\centering\fbox{\parbox[c][0.28\textheight][c]{0.84\textwidth}{\centering Placeholder: upload figures/figureS3.pdf}}\caption{Dendrogram from the hierarchical clustering used to identify the 18 bot types.}\label{fig:dendrogram}\end{figure}}\begin{figure}[t]
\centering
\includegraphics[width=\textwidth,keepaspectratio]{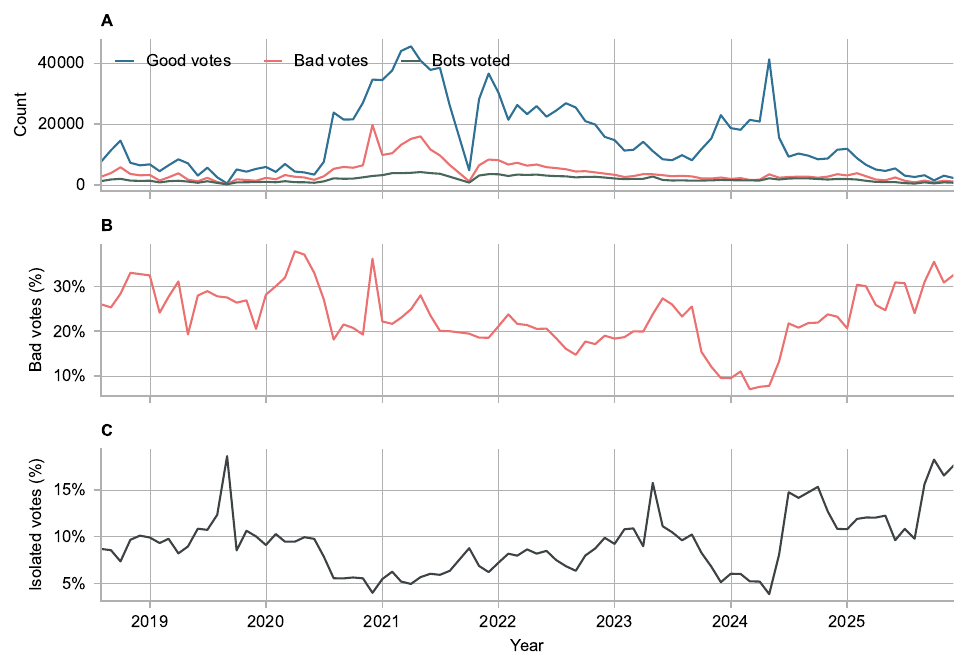}
\caption{The vote data from Botrank used to identify bots. A) Monthly voting activity. B) Monthly proportion of ``bad bot'' votes. C) Monthly proportion of votes for bots with one vote only.}
\label{fig:annual-composition}
\end{figure}

\clearpage
\renewcommand{\arraystretch}{1.3}
\begin{longtable}{@{}>{\RaggedRight\arraybackslash}p{0.26\textwidth}p{0.69\textwidth}@{}}
\caption{Description of the 33 features mined and the 23 most informative features (shown in bold) used in the clustering algorithm.}
\label{tab:features}\\
\toprule
\textbf{\textbf{Feature}} & \textbf{\textbf{Description}} \\
\midrule
\textbf{\textit{Temporal features}} &  \\
\textbf{Mean log inter-post time} & The average time interval in $\log_{10}$ seconds between consecutive posts. It captures the frequency of activity. \\
Burstiness & The extent to which activity is unevenly distributed over time, indicating whether posts occur in steady intervals or in concentrated bursts followed by periods of inactivity. $B = (sd - mean) / (sd + mean)$ for inter-arrival times; varies from regular ($-1$), through random ($0$), to bursty ($1$). \\
\textbf{Hourly entropy} & The degree to which posting activity is distributed across different hours of the day: lowest ($0$) for once-per-day scheduled tasks, lower for human-like circadian patterns, and highest ($1$) for posting equally across every hour of the day. \\
Mean log response time & The delay with which a bot responds in conversations, measured as the mean of $\log_{10}$-transformed time gap in seconds between parent's and bot's posts; set to $\log_{10}{(604800+1)}$ for submissions. \\
\textbf{Variance log response time} & The standard deviation of the delay with which a bot responds in conversations, measured as $\log_{10}{(x+1)}$ of the time gap in seconds between parent's and bot's posts. \\
\midrule
\textbf{\textit{Community features}} &  \\
\textbf{Log subreddits} & Thelog10number of distinct subreddits in which a bot is active, reflecting whether the bot operates broadly across communities or within a limited set of contexts. \\
\textbf{Subreddit specialization} & The distribution of activity across subreddits. We use the Herfindahl--Hirschman Index = sum of squared market share (originally measures market concentration) to measure how much the user's activity is dominated by a set of subreddits; $1$ if activity is dominated by one subreddit, low if activity is equally distributed among 10+ subreddits). \\
Moderation & The proportion of a bot's posts that are marked as distinguished, indicating involvement in moderation or official roles within communities. \\
Submission frequency & The extent to which the bot contributes via submissions rather than comments, measured as the share of all contributions that are submissions. \\
Post uniqueness & The degree to which a bot reuses identical posts, operationalized as the proportion of posts that are unique. \\
\textbf{Explicit trigger} & The proportion of comments in response to parent posts that mention the bot's username with ``u/botname'' or contain an explicit command cue with ``!command''. \\
\textbf{Similarity to parent} & The mean lexical similarity between the parent posts and the bot's replies, measured using Jaccard similarity. \\
\midrule
\textbf{\textit{Linguistic features}} &  \\
Mean log post length & The average length of posts, measured as $\log_{10}{(tokens + 1)}$, capturing whether the bot produced short or long-form content. \\
Variance log post length & The extent to which post lengths fluctuate relative to their average length, operationalized as the standard deviation of token length of posts. \\
\textbf{Log lexicon size} & The breadth of language across posts, measured as the $\log_{10}{(unique\ tokens + 1)}$. \\
\textbf{Lexical diversity} & The extent to which language in the bot's posts is diverse. Herdan's C = $\log(unique\ tokens + 1)$ / $\log{(total\ tokens + 1)}$. \\
Text self-similarity & The degree to which posts remain semantically similar to each other, operationalized as the mean centroid cosine similarity: for each bot, we take a sample of 500 contributions, each of which is represented as a vector of TF-IDF weights; we estimate the average cosine similarity between the contributions and their cluster's ``centroid'' (average vector) to evaluate cluster coherence, indicating how well documents within a cluster align with the central theme of that cluster; values closer to 1 suggest strong thematic similarity. \\
\textbf{Sentiment} & The average emotional tone of posts, measured using the VADER sentiment analysis tool. The compound score is used to capture overall sentiment, ranging from $-1$ (negative) to $1$ (positive). \\
Link frequency & The share of posts containing one or more URL links. \\
Expressiveness & How strongly posts signal emotion or emphasis through writing style. Computed by standardizing and averaging six rate-based cues, including emoji use, exclamation marks, question marks, ellipsis marks, repeated punctuation, and all caps. \\
\midrule
\textbf{\textit{Semantic features}} &  \\
\textbf{Adult/NSFW} & The proportion of a bot's posts contain explicit sexual material or clearly adult-oriented language and themes. \\
\textbf{Entertainment \& Fiction} & The proportion of a bot's posts engage with media and storytelling, including TV/films, books, comics/manga, anime, music, celebrities, fandom references, quotes, and lyrics. \\
\textbf{Gaming} & The proportion of a bot's posts are centered on video games and gaming communities, including game mechanics, in-game items, builds, matchmaking, mods, and game-specific alerts or reference information. \\
\textbf{Health \& Lifestyle} & The proportion of a bot's posts relate to everyday well-being and practical living, including health/medical, fitness, food, pets, safety, routines, and seasonal community threads. \\
\textbf{Memes \& Internet Culture} & The proportion of a bot's posts reflect internet-native humor and vernacular, including copypasta, in-jokes, catchphrases, shitposting, novelty call-and-response, and performative interaction styles. \\
\textbf{Platform Governance \& Moderation} & The proportion of a bot's posts are associated with community management functions, such as rule enforcement, content removal, behavioral guidance, etc. \\
\textbf{Politics and Public Affairs} & The proportion of a bot's posts address political actors and institutions, elections, policy debates, civic/public issues, and geopolitical events; religion is included here when it appears as public discourse or political identity. \\
\textbf{Relationships \& Social Life} & The proportion of a bot's posts provide interpersonal or emotional support, including dating/relationships, friendship, family, gratitude/hug/support messages, and social connection or matching functions. \\
\textbf{Shopping \& Consumer Products} & The proportion of a bot's posts are about consumer purchasing, including products, deals, pricing, reviews, authenticity/fake-product detection, shopping platforms, and purchase logistics. \\
\textbf{Sports} & The proportion of a bot's posts concern sports teams, leagues, matches, athletes, statistics, recruiting, and sport-specific updates. \\
\textbf{Technology \& Programming} & The proportion of a bot's posts focus on software, coding, computing infrastructure, developer tools, and technical troubleshooting. \\
\textbf{Utility Roles} & The proportion of a bot's posts are domain-neutral automation output, such as link mirrors, converters, counters, spell/grammar correction, reminders, notifications, and generic helper templates not tied to a specific semantic domain. \\
\textbf{Finance \& Economy} & The proportion of a bot's posts concern money and economic activity, including markets, investing, crypto/tokens, personal finance, labor/economic framing, and finance-community discourse. \\
\bottomrule
\end{longtable}

\end{document}